\documentclass[aps, twocolumn]{revtex4}
\usepackage{graphicx}
\usepackage{psfrag}
\usepackage{color}
\usepackage{latexsym}

\def\d {{\rm d}}
\renewcommand{\vec}[1]{{\bf #1}}

\begin{document}
%\bibliographystyle{prsty}

%\title{Mechanical Noise and Plastic Response of Athermally Sheared Amorphous Solids}
\title{Rate-Dependent Avalanche Size in Athermally Sheared Amorphous Solids}
\author{Ana\"el Lema\^{\i}tre$^{(1)}$}
\author{Christiane Caroli$^{(2)}$}
\affiliation{$^{(1)}$ Universit\'e Paris Est -- Institut Navier,
2 all\'ee Kepler, 77420 Champs-sur-Marne, France}
\affiliation{$^{(2)}$ INSP, Universit\'e Pierre et Marie Curie-Paris 6, 
CNRS, UMR 7588, 140 rue de Lourmel, 75015 Paris, France}

\date{\today}

\begin{abstract}
We perform an extensive numerical study of avalanche behavior
in a 2D LJ glass at $T=0$, sheared at finite strain rates $\dot\gamma$.
From the finite size analysis of stress fluctuations and of transverse diffusion
we show that flip-flip correlations remain relevant at all realistic strain rates.
We predict that the avalanche size scales as $\dot\gamma^{-1/d}$, with $d$ the space dimension.
%(i) even at very high strain rates, plasticity remains due to local flip events
%producing long-range elastic field with measurable effects.
%(ii) It is only at irrealisticaly high $\dot\gamma$ that correlations between flips become negligible.
%(iii) As $\dot\gamma$ decreases, correlation effects become increasingly important and the dynamics continuously 
%reaches its QS behavior below a system-size dependent cross-over $\dot\gamma_c$
%(iv) Above $\dot\gamma_c$, the length-scale of flip correlations (avalanche size) scales as $\dot\gamma^{-1/d}$, with $d$ the space dimension.
\end{abstract}

\maketitle

Considerable efforts have been spent in recent years to derive constitutive laws for plasticity in amorphous media
from a realistic description of the elementary mechanisms of dissipation. It is now agreed that, 
in these disordered systems, plasticity involves ``shear transformations'', i.e. irreversible rearrangements (or flips) 
of small clusters of (a few tens of) particles. 
By analogy with Eshelby transformations~\cite{Eshelby1957}, Argon and Bulatov~\cite{BulatovArgon1994a} 
inferred that such local rearrangements should generate
long-range elastic fields: hence, since each flip alters the strain field in its surroundings, the flowing system is 
submitted to a self-generated dynamical noise. 
From these premises, theories have developed along two very distinct lines.

(i) Several mesoscopic models~\cite{BulatovArgon1994a,BaretVandembroucqRoux2002,PicardAjdariLequeuxBocquet2005} 
explicitly incorporate long-range elastic interactions, but introduce
many phenomenological parameters, in order to take into account effects such as: thermal activation,
short-range disorder, flip duration,... which makes it quite difficult to test their assumptions
and evaluate their parameters.

(ii) Mean-field theories--STZ~\cite{FalkLanger1998} and SGR~\cite{Sollich1998}--rely upon the assumption 
that flips can be viewed as uncorrelated, local events 
activated by an effective thermal noise. These models, which do not specify how their effective temperature relates to
the dynamical noise, clearly overlook the possibility that elastic couplings 
may give rise to correlations between flips.
Indeed, it is well-known from studies on driven pinned systems--such as wetting lines or magnetic walls--that 
such long-range interactions in general give rise to avalanches (see~\cite{Colaiori2008} and Refs. therein). 

A first investigation of correlation effects has been performed by Maloney and Lemaitre 
using quasi-static (QS) simulations of sheared 2D glasses~\cite{MaloneyLemaitre2004a}. 
They found that, in this vanishing shear-rate regime, flips are not independent, random, events, 
but organize into avalanches, the size of which scales roughly as the linear size $L$ of the 
system. The existence of such avalanches, recently confirmed by Lerner and Procaccia~\footnote{
E. Lerner and I. Procaccia, cond-mat/0901.3477 (2009), confirm that avalanche size scales as $L^\kappa$, 
but find an exponent $\kappa={1.37}$. 
This discrepancy with~\cite{MaloneyLemaitre2004a} might stem
from algorithmic differences between QS protocols.},
strongly supports the view that correlation effects are essential under shear.
However, critiques have been raised, on the grounds that the QS limit could be singular, and that observations
made in this regime would not carry over to finite strain rates $\dot\gamma$.
Indeed, under QS conditions, the duration of flips and acoustic propagation delays are irrelevant 
on the shearing time scale $\dot\gamma^{-1}\to\infty$.
This permits to separate unambiguously plastic events as stress or energy drops occurring
over a zero strain interval. The scaling with $L$ of the drop sizes then suffices
to demonstrate the existence of avalanches.
At finite $\dot\gamma$, such a separation can no longer be performed, and one must devise other methods
to characterize possible spatio-temporal correlations between flips.

In this Letter, we present extensive numerical simulation results on a 2D LJ glass at $T=0$, 
driven over a wide range of finite strain rates, for various system sizes. 
We rely on systematic analysis of stress fluctuations and of the self-diffusion coefficient in steady state,
helped by direct imaging of the velocity and strain fields.
A heuristic decomposition of the dynamical noise leads us to a successful scaling prediction for 
the $L$- and $\dot\gamma$-dependence of the diffusion coefficient.
We conclude that: (i) even at very high strain rates, plasticity remains due to local flip events
producing long-range elastic field with measurable effects.
(ii) It is only at irrealisticaly high $\dot\gamma$ that correlations between flips become negligible.
(iii) As $\dot\gamma$ decreases, correlation effects become increasingly important and the dynamics continuously 
reaches its QS behavior below a system-size dependent cross-over $\dot\gamma_c$.
On this basis, we propose that, above $\dot\gamma_c$, 
the length-scale of flip correlations (avalanche size) scales as $\dot\gamma^{-1/d}$, with $d$ the space dimension.

We use the same 2D binary LJ mixture as that of Ref.~\cite{MaloneyLemaitre2006}, and we work in standard reduced LJ units. 
Large~(L) and small~(S) particle radii and numbers are $R_L=0.5, R_S=0.3, N_L=N_S(1+\sqrt{5})/4$, and particles 
have equal masses $m=1$. These values ensure that no crystallization occurs. 
The packing fraction of our $L\times L$ systems is $\pi(N_LR_L^2+N_SR_S^2)/L^2=0.9$ 
as in~\cite{LemaitreCaroli2007}, with $L$ ranging from 10 to 160. 
%Our time unit $\tau_{\rm LJ}$ corresponds to a physical time $\mathcal{O}(10^{-13}{\rm s})$. 

In our glassy system, Rayleigh scattering results in preferential damping of short wavelength sound modes. In order to implement $T=0$ dynamics, we thus introduce dissipative interparticle forces of the form:
$$
{\bf f}_i^{\rm visc.}=\frac{m}{\tau}\,\sum_{j}\phi\left(\frac{r_{ij}}{R_i+R_j}\right)\,\left({\bf v}_j-{\bf v}_i\right)
$$
with ${\bf v}_{i}$ a particle velocity, and $\phi(x)$ a nearly flat, normalized, weight function vanishing at the LJ 
cut-off $x=2$. We choose $\phi(x)\propto1-2(x/2)^4+(x/2)^8$, which guarantees sufficient smoothness.
This form of dissipation, similar to that used by Maloney and Robbins,~\cite{MaloneyRobbins2008} guarantees overdamping of sound for wavelengths $\lambda<\lambda_c\sim\pi a^2/\tau c_s$ (with $c_s\simeq3.4$ the transverse sound speed in our system and $a\sim1$ a typical interparticle distance). We take $\tau = 0.2 \tau_{\rm LJ}$, so that $\lambda_c/a\sim 5$.

Simple shear deformation is imposed along the $x$ axis, using Lees-Edwards boundary conditions and SLLOD dynamics. The integration timestep is usually $dt=10^{-2}\,\tau_{\rm LJ}$, but smaller values are used when needed to resolve small strain steps for large values of $\dot\gamma$. All configurations are prepared by rapid quench from a random initial state. A constant shear rate $\dot\gamma$ is then applied and steady state is reached beyond typical strains $\gamma\simeq0.3$. All data presented here have been obtained for strains ranging from 1 to at least 4, which guarantees their relevance to steady state dynamics. Strain rates $\dot\gamma$ range between $5.10^{-5}$ and $10^{-2}$. 

%Although this corresponds to very large physical values ($\gtrsim5.10^8\,{\rm s}^{-1}$), the lower limit turns out to be sufficiently small to enable us to bridge satisfactorily between QS and dynamical results.

\begin{figure}[h]
\resizebox{0.115\textwidth}{!}{\includegraphics{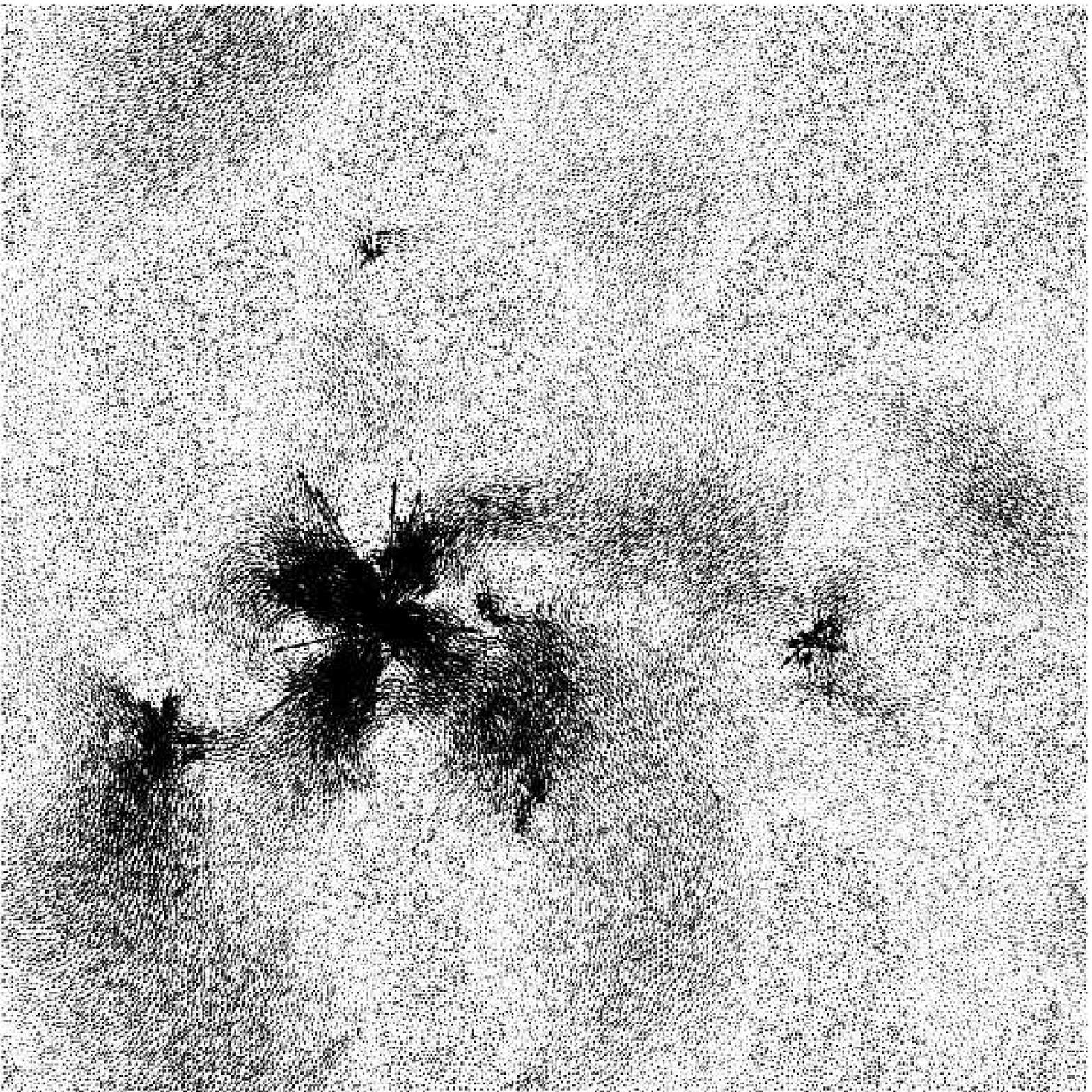}}
\resizebox{0.115\textwidth}{!}{\includegraphics{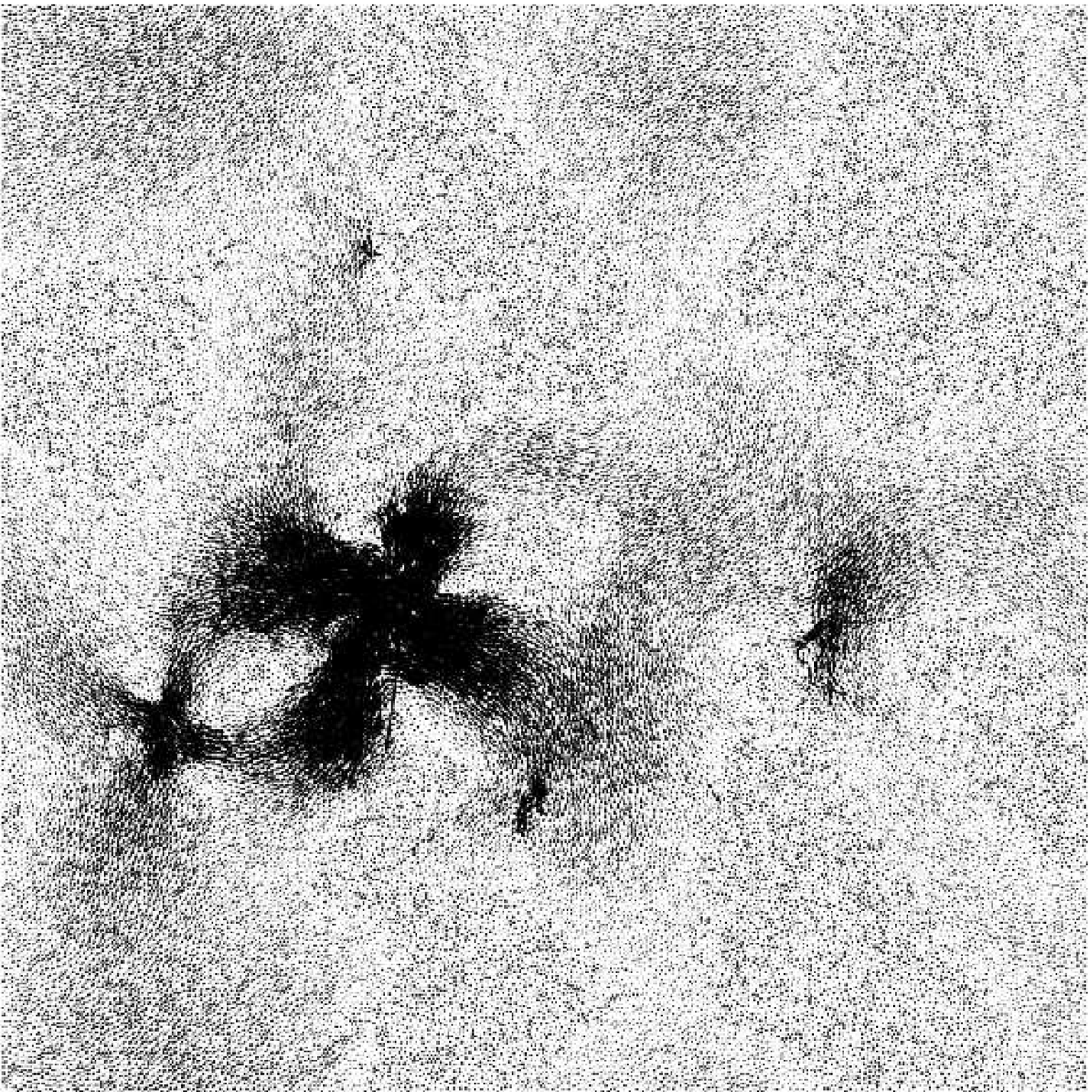}}
\resizebox{0.115\textwidth}{!}{\includegraphics{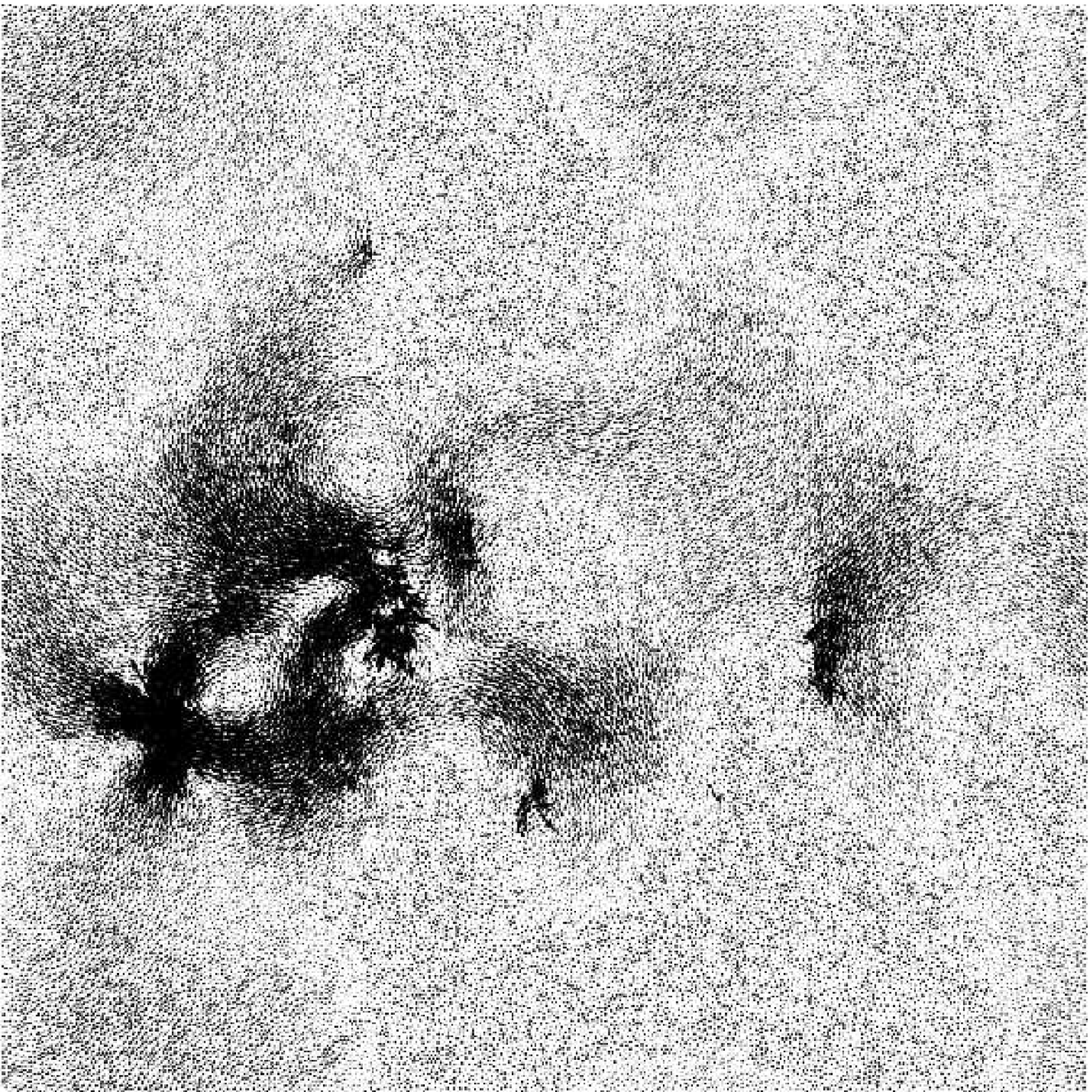}}
\resizebox{0.115\textwidth}{!}{\includegraphics{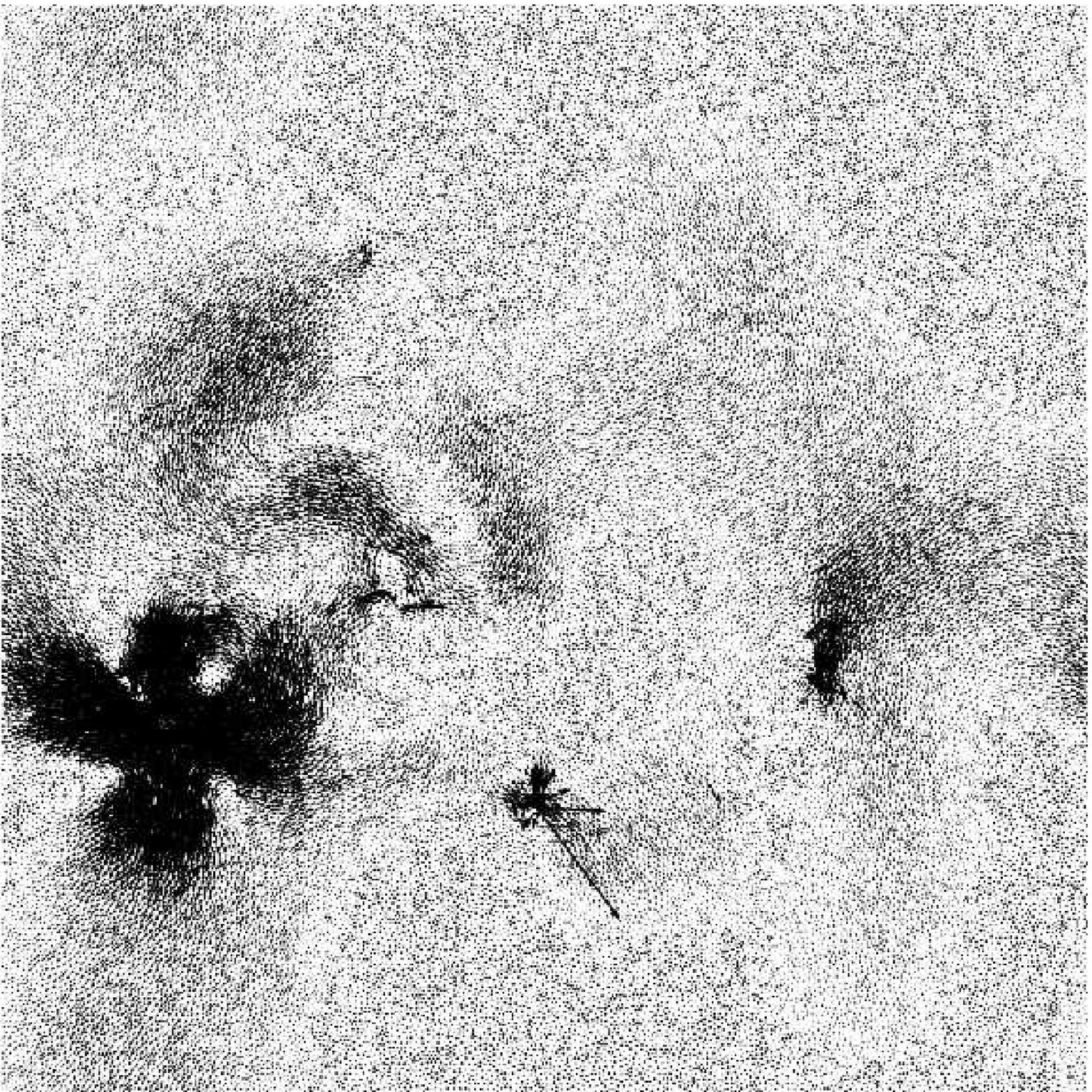}}
\caption{The instantaneous velocity field of a $160\times160$ system sheared at $\dot\gamma=5.10^{-5}$, at equally spaced times, with $\Delta t=2$, $\Delta\gamma=10^{-4}$. We see one zone which grows then disappears and triggers the flip of a nearby zone. A movie with the same parameters, covering a total time interval $\Delta t=20$, can be found in~\ref{movie}.}
\label{fig:quadrupoles}
\end{figure}

In order to access the qualitative features of the dynamics, we first inspect the instantaneous particle velocity field. Typical images (see Fig.~\ref{fig:quadrupoles}) exhibit highly localized regions of enhanced mobility, with roughly quadrupolar symmetry, which can be identified as ``weak'' zones close to their instability threshold. The emergence, growth, and subsequent decay of these quadrupolar structures, visible on time series and movies~\footnote{\label{movie} See movie in Supplementary Material.}, demonstrate that the plastic response results from the accumulation of flip events. This phenomenology is closely similar to that observed in QS simulations, yet it is also visible on movies that flip events trigger the emission of signals propagating out through the medium at velocities compatible with the transverse sound speed. This is a direct confirmation of the idea that shear-induced local rearrangements promote a long-range elastic field, i.e. can be viewed as Eshelby transformations.

%In the QS limit, the duration of flips and acoustic propagation delays are irrelevant 
%on the shearing time scale $\dot\gamma^{-1}\to\infty$.
%This permits to separate unambiguously plastic events as stress or energy drops occurring
%over a zero strain interval. The scaling of these events' sizes with $L$, then suffices
%to demonstrate that flips do not occur as uncorrelated random events, but as avalanches,
%most probably resulting from the presence of the long-range Eshelby fields. 
%At finite $\dot\gamma$, such a separation can no longer be performed, and one must devise other methods
%to characterize possible spatio-temporal correlations between flips.

\begin{figure}[h]
\resizebox{0.45\textwidth}{!}{\includegraphics{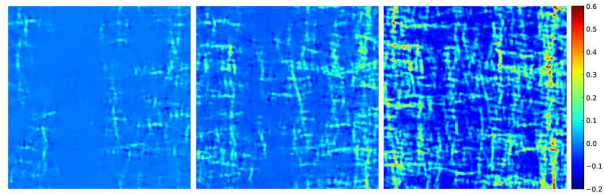}}
\caption{
The strain field of a $160\times160$ system sheared at $\dot\gamma=10^{-4}$, 
for growing strain intervals $\Delta\gamma = $1\%, 5\%, and 20\% (from left to right) 
from the same initial configuration.
} 
\label{fig:strain}
\end{figure}

A second qualitative piece of information can be obtained by constructing maps of 
the coarse-grained non-affine strain field $\epsilon_{xy}({\bf r}; \Delta\gamma)$ 
accumulated over moderate macroscopic strain intervals $\Delta\gamma$~\footnote{
We compute the non-affine strain field from a displacement field obtained by integrating 
over the time interval $\Delta t=\Delta\gamma/\dot\gamma$
the non-affine part of Goldhirsch's linear expression
(equation~(12) of~\cite{GoldhirschGoldenberg2002}).}.
A typical sequence of such maps is shown on Figure~\ref{fig:strain}, for a system 
of size $L=160$, at $\dot\gamma=10^{-4}$.
The existence of spatial correlations is obvious: flips gradually organize themselves into quasi-linear patterns
roughly aligned with the Bravais axes of the simulation cell. 
At this strain rate, an interval $\Delta\gamma=1$\% corresponds
to a time $\Delta t=100$, during which the acoustic signal propagates over a distance $\sim 34\sim L/5$. 
This is the maximal length-scale at which the information that a flip has occurred is transmitted through the elastic 
medium, hence it is an upper bound to the size of possible avalanches.
Indeed, it compares well with the size of the linear features seen on Figure~\ref{fig:strain}-(a). 
As $\Delta\gamma$ ($\Delta t$) increases, a criss-cross pattern, analogous to that observed by Maloney and 
Robbins,~\cite{MaloneyRobbins2008} gradually builds up due to both the preferential
triggering of flips along preexisting lines and the random emergence of new ones (see Figure~\ref{fig:strain}-(b-c)).
Clearly, avalanches persist at finite strain rates.

\begin{figure}[ht]
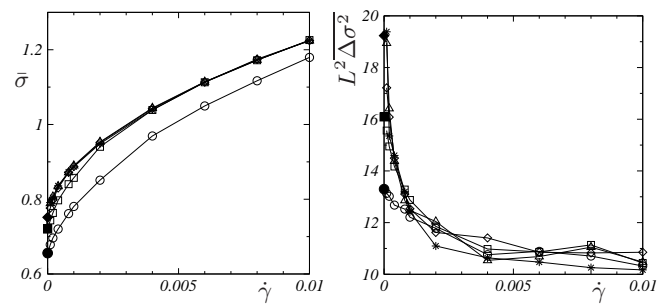

\psfrag{s}{{$\bar\sigma$}}
\psfrag{varsigma}{{$L^2\overline{\Delta\sigma^2}$}}
\psfrag{g}{{$\dot\gamma$}}
\includegraphics[width=0.23\textwidth]{sigma.eps}\hspace{2mm}
\includegraphics[width=0.23\textwidth]{varsigma.eps}
\caption{
For different sizes $L=10$ ({\Large $\circ$}), 20 ($\Box$), 40 ($\Diamond$), 80 ($\triangle$), 160 ($\star$):
macroscopic stress $\bar\sigma$ (left) and $L^2\overline{\Delta\sigma^2}$ (right)
versus strain rate $\dot\gamma$.
In both pictures QS values (for $L=10,20,40$) are indicated by full symbols.
}
\label{fig:stress}
\end{figure}

The question, however, remains to quantify the importance of these correlations. 
As a first characterization, we focus on the macroscopic stress $\sigma$ in steady state.
We compute its configurational average $\bar\sigma(\dot\gamma)$ and its fluctuations $\overline{\Delta\sigma^2}$,
for various $\dot\gamma$ and system sizes. 
As shown on Figure~\ref{fig:stress} (left), $\bar\sigma(\dot\gamma)$
quickly converges towards a size-independent limit (reached beyond $L\simeq40$),
for which an excellent empirical fit is: $\bar\sigma=0.74+4.87\,\sqrt{\dot\gamma}$.
We plot on Figure~\ref{fig:stress} (right) the product $L^2\,\overline{\Delta\sigma^2}$.
The data collapse for $\dot\gamma\gtrsim4\times10^{-3}$ shows that at the higher strain rates, 
stress fluctuations reasonably follow the law of large numbers, hence that 
the underlying dynamics is only weakly correlated. This collapse, however, becomes increasingly poorer
when $\dot\gamma$ decreases, indicating the growing importance of correlations. The spread culminates
in the small rate regime, where $L^2\,\overline{\Delta\sigma^2}$ extrapolates nicely to its QS values
(shown in full symbols), a strong hint that the underlying dynamics approaches continuously the QS behavior.

\begin{figure}[ht]
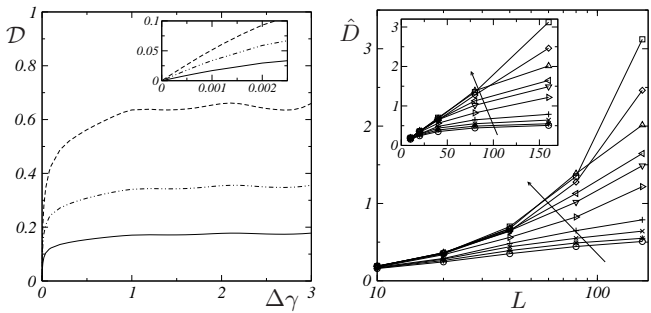

\psfrag{d}{{${\cal D}$}}
\psfrag{g}{{$\Delta\gamma$}}
\includegraphics[width=0.23\textwidth]{diffusion.eps}\hspace{2mm}
\psfrag{D}{{${\hat D}$}}
\psfrag{L}{{${L}$}}
\includegraphics[width=0.23\textwidth]{DofL.eps}
\caption{\label{fig:diffusion}
Left: Transverse displacement fluctuations, normalized as ${\cal D}=\langle \Delta y^2\rangle/2\Delta\gamma$,
as a function of $\Delta\gamma$, for $\dot\gamma=10^{-3}$ and
system sizes $L=20$ (solid line), 40 (dash-dotted), and 80 (dashed). 
Right: Their plateau value, $\hat D$ versus $L$ for strain rates:
$\dot\gamma=10^{-4}$, $2.10^{-4}$, $4.10^{-4}$, $8.10^{-4}$, $10^{-3}$, $2.10^{-3}$,
$4.10^{-3}$, $6.10^{-3}$, $8.10^{-3}$, $10^{-2}$. Arrows indicate decreasing $\dot\gamma$.
QS values, shown for sizes 10, 20, and 40, are
indistinguishable from lowest strain rates values.
}
\end{figure}

However, the fluctuations of such a macroscopic quantity only provide very global information about the dynamics.
To better qualify the spatio-temporal correlations, we turn to another observable which carries 
more microscopic information, namely the transverse (self)-diffusion coefficient. It is obtained 
from the space and ensemble average of the fluctuations $\overline{\Delta y^2}$ of the transverse displacements
$\Delta y_i=y_i(\gamma_0+\Delta\gamma)-y_i(\gamma_0)$.
To facilitate comparison with QS data, we introduce the quantity $\overline{\Delta y^2}/2\,\Delta\gamma$,
which is plotted versus $\Delta\gamma$ for different system sizes, on Figure~\ref{fig:diffusion}-(left).
At very short times (small $\Delta\gamma$'s), particle motion is ballistic, hence the curves
start with a finite slope (see insert). After a transient regime extending up to $\Delta\gamma\sim1$ they
reach a plateau value,
%~\footnote{The wiggles visible on the $L=80$ curve are due to a lack of sampling, as we checked on smaller systems, even though we have accumulated data from 50 configurations sheared up to 400\%.} 
$\hat D$, related to the usual diffusion coefficient $D$ as $D=\hat D\dot\gamma$. 
It is striking that, for a fixed $\dot\gamma$, $\hat D$ is strongly size-dependent.
We plot $\hat D$ as a function of $L$ on Figure~\ref{fig:diffusion}-(right), for various strain rates.
Although the size dependence is stronger near the QS limit where $\hat D(L)$ becomes nearly linear (see insert),
it persists up to the highest value $\dot\gamma=10^{-2}$, for which $\hat D$ is quasi-linear in $\ln L$.

It is surprising that $\hat D$ remains size-dependent for $\dot\gamma$ values at which stress fluctuations were shown
to obey the law of large numbers, an indication that plastic events are then uncorrelated.
The only way to reconcile these two observations is to assume that each event induces a long-range elastic field,
which is precisely the case of Eshelby-like transformations. In our simple shear geometry the relevant transformations
are purely deviatoric sources. In an infinite medium,
the expression of the displacement field due to such a source located at the origin
is~\cite{PicardAjdariLequeuxBocquet2004}:
$\vec u_{\rm E} = \frac{2a^2 \Delta\epsilon_0}{\pi}\,\frac{xy}{r^4}\,\vec r$, with $a$ the source size, 
and $a^2\,\Delta\epsilon_0$ its dipolar strength.
For $L\gg a$ the transverse displacement fluctuation due to a single flip
can then be computed to leading order as a space average:
$\langle u_{y,{\rm E}}^2\rangle\approx\frac{a^4\,\Delta\epsilon_0^2}{2\,\pi\,L^2}\,\ln(L/a)$.
Since each flip releases a macroscopic strain $\frac{a^2\,\Delta\epsilon_0^2}{L^2}$
(up to a prefactor of order 1 which characterizes the local zone structure)
the average number of flips, in steady state, occurring over a strain range $\Delta\gamma$ is: 
$N_{\rm f}(\Delta\gamma)=\frac{L^2\,\Delta\gamma}{a^2\,\Delta\epsilon_0}$. 
Assuming the flips uncorrelated, we thus obtain 
$\hat D = N_{\rm f}\,\langle{u_{y, \rm E}^2}\rangle/2\,\Delta\gamma\approx\frac{a^2\,\Delta\epsilon_0}{4\,\pi}\,\ln(L/a)$.
In light of this result, our data for $\dot\gamma=10^{-2}$ demonstrate that at high $\dot\gamma$ the dynamics
results from uncorrelated Eshelby flips. A fit of $\hat D(L)$ at $\dot\gamma=10^{-2}$ yields a value of order 1
for $a^2\,\Delta\epsilon_0$, which is reasonable since we expect $a\sim 4-5$, and $\Delta\epsilon_0\sim$ a few percents.

The growing departure from the $\ln L$ scaling at decreasing $\dot\gamma$ must then 
be due to the emergence of correlations between flips. 
Guided by the observation of quasi-linear avalanches in strain maps, we now 
try out a model in which diffusion results from a set of independent events, each of which is a linear avalanche
of length $\ell$, oriented at random along either the $x$ or $y$ axis. 
In such a picture, the existence of flip-flip correlations is embodied in the value of $\ell$.
We assume the linear density of flips $\nu$ in an avalanche to be a constant,
and compute the transverse displacement fluctuations due to a line of homogeneously distributed flips as:
$
\langle{\Delta y^2}\rangle_{\rm A}=\nu^2\,\int_0^\ell\int_0^\ell\d s\,\d s' C(\vec r_s-\vec r_{s'})$
with $C(\delta \vec r)=\langle{u_{y,\rm E}(\vec r)\,u_{y,\rm E}(\vec r+\delta \vec r)}\rangle
$. For $|\delta\vec r|\ll L$, to leading order,
$C(\delta \vec r)\approx \frac{a^4\,\Delta\epsilon_0^2}{2\,\pi\,L^2}\,\int_{|\delta\vec r|/L}^\infty\,\frac{\d q}{q}\,J_0(q)$,
whence 
$\langle{\Delta y^2}\rangle_{\rm A}\approx\frac{a^4\,\Delta\epsilon_0^2\,\nu^2}{2\,\pi}\,\frac{\ell^2}{L^2}\ln(L/\ell)$.
Since the average number of avalanches over an interval $\Delta\gamma$ is $N_{\rm A}=N_{\rm f}/\nu\,\ell$,
the resulting diffusion coefficient is:
\begin{equation}
\label{eq:D}
\hat D\approx\frac{a^2\,\Delta\epsilon_0}{4\,\pi}\,\nu\,\ell\,\ln(L/\ell)\quad .
\end{equation}
Of course, at this stage, $\ell$ is an unknown function of both $\dot\gamma$ and $L$.
About it we only know that (i) $\ell\sim a$ at large $\dot\gamma$, and (ii) $\ell\propto L$ in the QS limit.
Our model thus does capture the limiting, logarithmic and quasi-linear, scaling behaviors in these limits.

\begin{figure}[t]
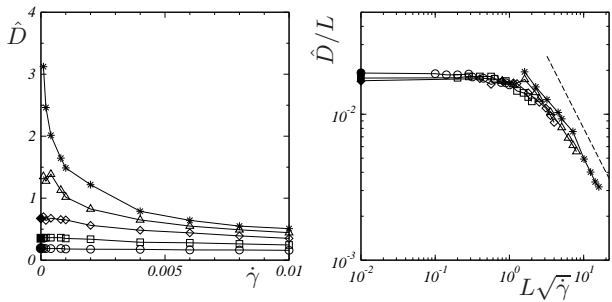

\psfrag{D}{{${\hat D}$}}
\psfrag{g}{{$\dot\gamma$}}
\includegraphics[width=0.22\textwidth]{Draw.eps}\hspace{2mm}
\psfrag{Dsc}{{${\hat D/L}$}}
\psfrag{gsc}{{${L\sqrt{\dot\gamma}}$}}
\includegraphics[width=0.22\textwidth]{scaling.eps}
\caption{
For sizes $L=10$ ({\Large $\circ$}), 20 ($\Box$), 40 ($\Diamond$), 80 ($\triangle$), 160 ($\star$):
Left: $\hat D$ vs $\dot\gamma$.
Right: $\hat D/L$ vs $L\,\sqrt{\dot\gamma}$. The dashed line has slope -1.
QS values for $L=10, 20, 40$ are shown as full symbols on the vertical axis. 
}
\label{fig:scaling}
\end{figure}

For further comparison with the data, we seek to complement this model with an estimate of the dependence 
of $\ell$ on $\dot\gamma$. Let us recall that each zone receives noise due to elastic signals propagating 
away from flips occurring 
in the whole system system at rate ${\cal R}=\frac{L^2\,\dot\gamma}{a^2\,\Delta\epsilon_0}$. 
Each signal carries directional information and gives rise to a stress jump
with rise time (or autocorrelation time) $\tau\sim \eta^{-1}\sim a/c_s$.
Given a distance $\ell$, we distinguish between: (i) signals originating from nearby sources (within $r < \ell$) 
occuring at rate ${\cal R}_\ell={\cal R}\ell^2/L^2$, and of amplitude 
$\Delta\sigma_0\gtrsim\mu\frac{a^2\,\Delta\epsilon_0}{\ell^2}$;
(ii) a background noise due to all other sources, of rate ${\cal R}'_\ell={\cal R}-{\cal R}_\ell$.
It is incoherent and isotropic, the sources being evenly distributed in space,
as soon as several far-field signals overlap at any time (${\cal R}'_\ell\gg \tau^{-1}$).
We assume $\ell \ll L$ so that ${\cal R}'_\ell\simeq {\cal R}$.
We then make the Ansatz that $\ell$ is the flip-flip correlation length if the near-field signals constitute 
a shot-noise which stands out of the incoherent background.
This entails two conditions:
(a) near-field signals must not overlap: ${\cal R}_\ell\lesssim\tau^{-1}$; 
(b) their amplitude $\Delta\sigma_0$ must be larger than the background stress fluctuations
accumulated during $\tau$:
$\overline{\Delta\sigma^2}\sim\dot\gamma\tau\frac{\mu^2\,a^2\,\Delta\epsilon_0}{\ell^2}$.
Both lead to a common estimate for the correlation length:
\begin{equation}
\label{eq:ell}
\ell \approx \sqrt{\frac{a^2\,\Delta\epsilon_0}{\dot\gamma\tau}}
\quad,
\end{equation}
which also guarantees the incoherence of the background (${\cal R}'_\ell\simeq {\cal R}\gg \tau^{-1}$). Of course, this expression is valid only for $\ell\ll L$, as $\ell$ should saturate to its QS plateau value $\sim L$, below a 
cross-over strain-rate 
$\dot\gamma_c\sim\frac{a^2\,\Delta\epsilon_0}{\tau\,L^2}$.
This argument predicts that the diffusion coefficient $\hat D$ should obey a scaling of the form:
$\frac{\hat D}{L}=f(L\,\sqrt{\dot\gamma})$, with $f(x)\sim 1/x$ for $x\gg1$. We test it using the 
$\hat D(\dot\gamma)$ data presented on Figure~\ref{fig:scaling}.
Not only is the collapse satisfactory, but the prediction for a -1 slope at high $\dot\gamma$ 
fits the data remarkably well (in view of the numerous simplifying assumptions of our model).
The cross-over occurs near $L\,\sqrt{\dot\gamma}\simeq 1$, which is compatible with our expression for $\dot\gamma_c$
since $\tau\simeq0.2$ and $a^2\,\Delta\epsilon_0$ was found to be of order 1 from the analysis 
of $\hat D(L)$ at $\dot\gamma=10^{-2}$.
These results nicely support our simple argument and validate the prediction, equation~(\ref{eq:ell}), 
for the strain-rate dependence of the avalanche size. 

As a side note, let us recall that it is common to estimate the departure from the yield stress as
$\bar\sigma(\dot\gamma)-\bar\sigma_y\approx \mu\,\tilde\tau\,\dot\gamma$, with $\tilde\tau$ a microscopic time.
If we take it to be an avalanche duration $\tau_A\sim\ell(\dot\gamma)/c_s$ we obtain
the right functional behavior, $\bar\sigma(\dot\gamma)\approx\bar\sigma_y+C\,\sqrt{\dot\gamma}$, 
with $C\approx\frac{\mu}{c_s}\,\frac{a^2\Delta\,\epsilon_0}{\tau}=13$, a quite decent guess compared to the value 
$\sim 5$ which fits our data.

%Note finally that, as the avalanche duration $\tau_A\sim\ell(\dot\gamma)/c_s$ we can tentatively evatuate $\bar\sigma(\dot\gamma)\approx\bar\sigma_y+C\,\sqrt{\dot\gamma}$, with $C\approx\frac{\mu}{c_s}\,\frac{a^2\Delta\,\epsilon_0}{\tau}=13$. Not only does this predict the right functional behavior, but the value of $C$ is decently compatible with the value $\sim 5$ which fits our data.

To sum up, we have studied plastic flow in a 2D amorphous system over a broad range of shear rates.
We find that the flip-flip correlations due to long-range elastic Eshelby fields, observed in the QS simulations,
persist under dynamic conditions. In other words, zone flips are not in general independent random events,
but occur as directional quasi-linear avalanches, whose average size $\ell$ depends on $\dot\gamma$.
Below a cross-over rate $\dot\gamma_c$, any finite size system presents a pseudo-QS regime in which $\ell$
plateaus at its QS value $\ell\sim L$, while all flow properties converge to their QS values. 
Beyond $\dot\gamma_c$, $\ell(\dot\gamma)$ roughly scales as $\sim 1/\sqrt{\dot\gamma}$.
Its cross-over value therefore diverges in the large system size limit.
It is only at very large strain rates, beyond at least $\dot\gamma\sim10^{-2}$, that correlations 
become negligible. This is the only limit in which the existing mean-field models,
based upon independent flips activated by an incoherent noise -- modeled as an 
effective temperature -- are justified.

Of course, our results pertain to 2D systems and call for 3D numerical simulations of diffusion under shear.
Pending such studies, we tentatively carry over our Ansatz to 3D. It predicts that avalanches should extend 
over a length-scale $\ell(\dot\gamma)\sim a\,\left({\Delta\epsilon_0}/{\dot\gamma\tau}\right)^{1/3}$. 
This permits to estimate the order of magnitude of avalanche extent for particular classes of materials.
For instance, for atomic or molecular glasses, $\tau_{\rm LJ}\approx 10^{-13}{\rm s}$ 
and a typical maximum strain rate is $\dot\gamma\lesssim 1{\rm s}^{-1}$. Using $\Delta\epsilon_0\sim$5\% and
$a\sim 5$ LJ unit length, $\approx 1{\rm nm}$, we obtain a minimum value for $\ell$ in the 10$\mu{\rm m}$ range.
The fully decorrelated regime $\ell\to a$ would then be attained for completely irrealistically large strain rates 
$\dot\gamma\gtrsim 10^{9-10}{\rm s}^{-1}$.
% Based on this estimates..
We thus claim that, in these systems, avalanches should always be relevant under usual experimental conditions.

\end{document}